# Automatic Library Generation for Modular Polynomial Multiplication

Lingchuan Meng

Drexel University, Philadelphia PA 19104, USA
lm433@cs.drexel.edu

**Abstract.** Polynomial multiplication is a key algorithm underlying computer algebra systems (CAS) and its efficient implementation is crucial for the performance of CAS. In this paper we design and implement algorithms for polynomial multiplication using approaches based the fast Fourier transform (FFT) and the truncated Fourier transform (TFT). We improve on the state-of-the-art in both theoretical and practical performance. The SPIRAL library generation system is extended and used to automatically generate and tune the performance of a polynomial multiplication library that is optimized for memory hierarchy, vectorization and multi-threading, using new and existing algorithms. The performance tuning has been aided by the use of automation where many code choices are generated and intelligent search is utilized to find the "best" implementation on a given architecture. The performance of autotuned implementations is comparable to, and in some cases better than, the best hand-tuned code.

**Keywords:** polynomial arithmetic, convolution, fast Fourier transform, truncated Fourier transform, code generation and optimization

## 1 Introduction

Polynomial arithmetic is a key component of symbolic computation, scientific computing and cryptography. CAS and certain cryptographic algorithms perform computations using exact arithmetic with integers, rational numbers, polynomials, rational functions, algebraic numbers and functions, and finite fields. Many algorithms for computing in those domains use modular techniques. Moreover, multiplication with integer can be based on fast modular polynomial multiplication using the three primes algorithm or other related approaches. Furthermore, multivariate polynomial multiplication can be reduced to univariate polynomial multiplication using evaluation and interpolation. This justifies our focus on fast modular polynomial multiplication.

High performance implementations of polynomial arithmetic have become increasingly difficult to achieve on modern processors, despite the abundance of fast algorithms and the peak performance they provide. Traditionally, polynomial arithmetic libraries are manually developed and tuned with limited vectorization and multi-threading. This requires the programmers to have extensive domain knowledge and programming skills, and explore by hand many tunable parameters in addition to a large number of algorithms and implementation choices. Furthermore, a library tuned to a specific platform is in general not portable, which means that performance-critical libraries or subroutines must be reoptimized or reimplemented for new platforms.

Automatic code generation and optimization systems have been proved to be able to automatically produce implementations comparable to, and in some cases better than, the best hand-tuned codes. Examples include ATLAS [3, 41] and FLAME [13] for dense linear algebra, SPARSITY [17] for sparse linear algebra, FFLAS/FFPACK [4, 5] for finite field linear algebra, and FFTW [7, 8] for the fast Fourier transform (FFT) and SPIRAL [18, 33, 40, 32] for more general DSP algorithms and library generation. Autotuning-based systems employ various techniques to produce fast code for a platform-specific architecture and instruction set with minimal human intervention. These techniques incorporate domain-specific knowledge and architecture features to provide large space of implementation candidates, and utilize empirical benchmarking, search and learning to achieve optimal performance.

Despite the lack of autotuning for exact polynomial arithmetic, there has been extensive work on fast algorithms and hand-tuned high performance implementations. Algorithms and implementations for polynomial arithmetic have been developed with reduced arithmetic complexity, better adaptation to parallel architectures and improved space efficiency [9, 21]. In [36, 37], the author introduced a new variation of Fourier transform to be used by convolution, aiming to reduce arithmetic cost and avoid zero-padding. Then [14] improved the original radix-2 TFT/ITFT algorithm by optimizing for cache locality. Recently, [30] introduced parallel algorithms for sparse polynomial multiplication using heaps, and [11, 16] focused on sparse interpolation for multivariate polynomial and for over finite fields, respectively. [2] compared Chinese Remainder Theorem-based convolution algorithms with FFT-based algorithms, and presented a framework for convolution algorithm derivation. On top of transforms and convolutions, [15, 38] developed fast algorithm for univariate and multivariate polynomial multiplication. [19] is among the first publications to investigate algorithms for sparse polynomial arithmetic.

The `modpn` library[20] has been integrated into Maple, providing hand-optimized low-level routines implementing FFT-based fast algorithms for multivariate polynomial computations over finite fields, in support of higher-level code. The implementation techniques employed in `modpn` are often platform-dependent, since cache size, associativity properties and register sets have a significant impact and are platform specific. [39] also contains hand-optimized implementations of modular arithmetic in its standard libraries for large numbers, polynomials, etc. based on FFT and other fast algorithms.

Recent work has enabled the automatic generation of high performance libraries for the underlying transforms used by the modular polynomial multiplication. [29, 28] were the first to present the generation of fixed-size small and medium modular FFTs with efficient vectorized modular arithmetic. The scope of library generation was then extended by [24] to support the autotuning for general-size parallel modular FFT library.[25, 27] further extended the TFT/ITFT algorithms to support general-radix factorization and parallelization, leading to performance improvement compared to "stair-case phenomenon" found in the power-of-two modular FFT libraries. Recently, [26, 23] explained the approaches to automatically generate modular polynomial multiplication library based on the underlying transforms.

**Contribution.** This paper provides an automatically generated and optimized library for modular polynomial multiplication. The vectorized and multithreaded library is automatically generated and optimized from high level specification. The autotuning is done through extending and using SPIRAL. General-radix and parallel algorithms for modular FFT, TFT and ITFT have been exploited via the Convolution theorem, in order to better adapt to the memory hierarchy, vectorization, and parallelization. The TFT and ITFT based implementations smooth performance between powers of two compared to state-of-the-art power-of-two FFT performance. This paper is based on the content of the recent thesis[22].

## 2 Background

In this section, we first focus on the definitions of important polynomial arithmetic operations. The definitions of the underlying transforms and their fast algorithms can be found in [24] for modular FFT, and in [25, 27] for truncated Fourier transforms.

We then describe the code generation mechanism, including a domain specific language called SPL with extensions including the $\sum$-SPL and the OL, and a library generator called SPIRAL which automatically rewrites and optimizes the SPL expressions and generate vectorized and parallel high performance libraries.

### 2.1 Polynomial Multiplication

We define univariate polynomial and polynomial multiplication. Then, we review the convolutions that are crucial in the fast algorithms for polynomial arithmetic.

**Definition 1.** *Let $R$ be a commutative ring, such as $\mathbb{Z}$, a univariate polynomial $a \in R[x]$ in $x$ is a finite sequence $(a_0, \ldots, a_n)$ of elements of $R$ (the coefficients of $a$), for some $n \in \mathbb{N}$, and we write it as*

$$a = a_n x^n + a_{n-1} x^{n-1} + \cdots + a_1 x + a_0 = \sum_{0 \le i \le n} a_i x^i. \tag{1}$$

In practice, dense and sparse polynomials require different representations in data structure. For dense polynomials, we can represent $a$ by an array who $i$th element is $a_i$. This assumes that we already have a way of representing coefficients from $R$. The length of this representation is $n + 1$.

Next, we define the univariate polynomial multiplication.

**Definition 2.** *Let* $a = \sum_{0 \le i \le n} a_i x^i$ *and* $b = \sum_{0 \le i \le m} b_i x^i$*, the polynomial product* $c = a \cdot b$ *is defined as* $\sum_{0 \le k \le m+n} c_k x^k$*, where the coefficients are*

$$c_k = \sum_{\substack{0 \le i \le n, 0 \le j \le m \\ i+j=k}} a_i b_j, \quad \textit{for } 0 \le k \le m+n. \tag{2}$$

The naive implementation of polynomial multiplication has an $\mathcal{O}(n^2)$ complexity. The Karatsuba algorithm reduces the cost to $\mathcal{O}(n^{1.59})$. The FFT introduced earlier can be used in convolutions to further enable fast algorithms with a complexity of $\mathcal{O}(n \log n)$.

The convolution is at the core of our polynomial multiplication library. Beyond polynomial arithmetic, convolution also has applications in signal processing and efficient computation of large integer multiplication and prime length Fourier transforms. Next, we present the definitions of linear and circular convolutions, and interpret them from different perspectives to show the connection between convolutions and the polynomial multiplication.

Both linear and circular convolutions can be viewed from three different perspectives: (1) as a sum, (2) as a polynomial product and, (3) a matrix operation. As a result, polynomial algebra can be used to derive algorithms and the corresponding matrix algebra can be used to manipulate and implement algorithms.

**Definition 3.** *Let* $u = (u_0, \ldots, u_{M-1})$ *and* $v = (v_0, \ldots, v_{N-1})$*. The linear convolution* $u * v$ *is defined as*

$$(u * v)_i = \sum_{k=0}^{N-1} u_{i-k} v_k, \quad 0 \le i < M + N \tag{3}$$

If $u$ and $v$ are viewed as the coefficient vectors of polynomials, i.e.,

$$u(x) = \sum_{i=0}^{M-1} u_i x^i, \quad v(x) = \sum_{j=0}^{N-1} v_j x^j,$$

then the linear convolution $u * v$ is equivalent to polynomial multiplication $u(x)v(x)$.

The sum form of linear convolution is also equivalent to the following matrix vector multiplication.

$$u * v = \begin{bmatrix} u_0 & & & \\ u_1 & u_0 & & \\ \vdots & u_1 & \ddots & \\ u_{M-1} & \vdots & \ddots & u_0 \\ & u_{M-1} & & u_1 \\ & & \ddots & \vdots \\ & & & u_{M-1} \end{bmatrix} \cdot v \tag{4}$$

The circular convolution of two vectors of size $N$ is obtained from linear convolution by reducing $i-k$ and $k$ in 3 modulo $N$.

**Definition 4.** *Let $u = (u_0, \ldots, u_{N-1})$ and $v = (v_0, \ldots, v_{N-1})$. The circular convolution $u \circledast v$ is defined as*

$$(u \circledast v)_i = \sum_{k=0}^{N-1} u_k v_{(i-k) \mod N}, \quad 0 \leq i < N. \tag{5}$$

Similar to the polynomial and matrix perspectives of linear convolution, the circular convolution can be obtained by multiplying polynomials $u(x)$ and $v(x)$ and taking the remainder modulo $x^N - 1$. In terms of matrix algebra, circular convolution can be interpreted as the product of a *circulant matrix* $Circ_N(u)$ times $v$, where all columns of the matrix are obtained by cyclically rotating the first column.

$$u \circledast v = \begin{bmatrix} u_0 & u_{N-1} & u_{N-2} & \cdots & u_1 \\ u_1 & u_0 & u_{N-1} & \cdots & u_2 \\ \vdots & \ddots & \ddots & \ddots & \vdots \\ u_{N-2} & \cdots & u_1 & u_0 & u_{N-1} \\ u_{N-1} & u_{N-2} & \cdots & u_1 & u_0 \end{bmatrix} \cdot v \tag{6}$$

The convolution theorem introduced in Section 3 leads to fast convolution algorithms based on the aforementioned linear transforms, where two forward and one inverse transforms are required. The OL extension reviewed in the next section can express the fast algorithms in the declarative representation, similar to the fast algorithms for transform being expressed in the SPL[24, 25, 27].

### 2.2 Code Generation Mechanism

**SPIRAL** In this paper, we use, extend and enhance the SPIRAL library generator for library generation and performance tuning. SPIRAL is a library generator that can generate stand-alone programs for DSP algorithms and numeric kernels. In contrast to other work in autotuning, SPIRAL uses internally the domain-specific language SPL to express divide-and conquer algorithms for transforms as breakdown rules. For a user-specified transform and transform size, SPIRAL applies these rules to generate different algorithms, represented in SPL. Then $\sum$-SPL makes explicit the description of loops and index mappings. The $\sum$-SPL representation of algorithms are then processed by the built-in rewriting systems for performing difficult optimizations such as parallelization, vectorization, and loop merging automatically.

The optimizations are performed at a high abstraction level in order to overcome known compiler limitations. The unparsers translate these algorithms into executable code. Based on the generated code, a search engine uses the dynamic programming technique at generation time or run-time to explore different choices of algorithms to find the best match to the computing platform. The performance-based search is chosen over any high-level derivation as it can precisely predict the practical performance of the algorithms. The library generation workflow in SPIRAL is illustrated in Fig. 1. Extensive details and examples can be found in [40].

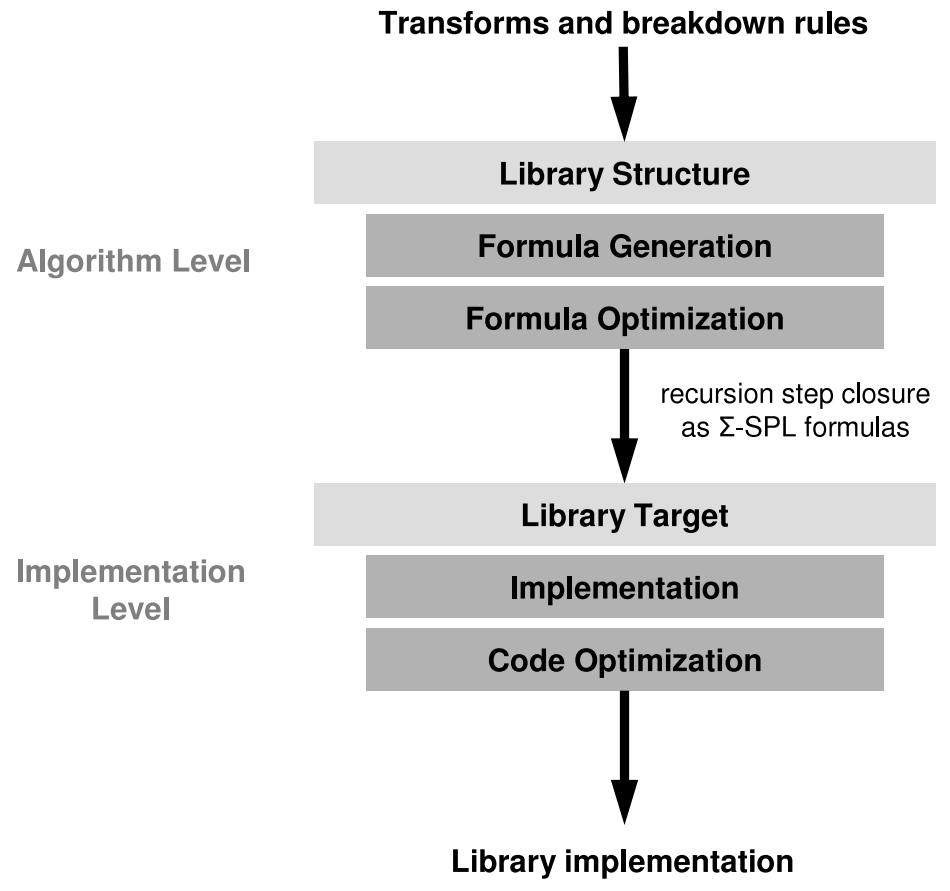

**Fig. 1.** Library generation in SPIRAL

New transforms can be added by introducing new symbols and their definitions, and new algorithms can be generated by adding new rules. SPIRAL was developed for floating point and fixed point computation; however, many of the transforms and algorithms carry over to finite fields. For example, the DFT of size $n$ is defined when there is a primitive $n$th root of unity and many factorizations of the DFT matrix depend only on properties of primitive $n$th roots. In this case, the same machinery in SPIRAL can be used for generating and optimizing modular transforms. Details of the generation of modular FFT library can be found in [24]. The more involved TFT and ITFT can also be automatically generated and optimized, as shown in [25, 27].

For the domain specific language, we focus on the OL which is used to express the fast algorithms for modular polynomial multiplication. More details of the SPL and $\sum$-SPL can be found with examples in [22].

**SPL** is a high-level domain-specific language that expresses recursive DSP algorithms at a high abstraction level as sparse matrix factorization. SPL expressions can be derived and optimized by the rewriting systems in SPIRAL to adapt to target computing platforms. The detailed examples of SPL can be found in [22].

**$\sum$-SPL** was introduced in [40] as an extension of the SPL designed to solve the formal loop merging problem and to enable complex data manipulation. The transition of SPL $\rightarrow \sum$-SPL is usually automated, provided that the algorithms can be completely described by SPL.

**OL**, a framework for automatic generation of fast numerical kernels, is recently developed as an extension to the SPL and integrated into SPIRAL. OL provides the structure to extend SPIRAL beyond the transform domain. The control flow of many kernels is data-independent, which allows OL to cast their algorithms as operator expressions. In this paper, the multi-linear operations like the linear and circular convolutions require using OL.

The three main components of OL framework are:

- the operator language (OL) that describes the kernel algorithms;
- the hardware tags that describe architecture features; and
- the tagged OL, a common abstraction of architecture and algorithms.

The basic building blocks of OL are operators that are $n$-ary functions on vectors. The arity of an operator is a 2-tuple $(r, s)$, meaning that the operator consumes $r$ vectors and produces $s$ vectors. Note that the linear transforms including modular FFT, TFT, and ITFT all have the same arity $(1, 1)$. It is also worth noting that the matrices are viewed as vectors stored linearized in row major order. Therefore, the permutation operator $L_r^{rs}$ that transposes an $r \times s$ matrix is of arity $(1, 1)$.

The basic operators can be combined into OL formulae by the high-order operators, as the parameters provided to the functions on operators. For example, the composition, denoted by $\circ$, is a generalization of the matrix multiplication in SPL that composes adjacent linear transforms. For instance,

$$L_r^{rs} \circ P_{rs}$$

is an arity $(2, 1)$ operator formula that first multiplies point-wise two matrices and then transposes the result.

The cross product of two operators, denoted by $\times$, is used in the convolution formula where two forward transforms are applied to two input vectors. It applies the first operator to the first input set and the second operator to the second input sect, and then combines the outputs. For example,

$$L_r^{rs} \times P_{rs}$$

is an arity $(3, 2)$ operator formula that transposes the its first input, and multiplies pointwise the second and third inputs, producing two output vectors.

The most important SPL operator, the tensor product $\otimes$, is also the most important higher order operator in OL, after it is formally extended. The generalization of the tensor product from SPL to OL is omitted here, as it is beyond the scope of this paper.

Recursive algorithms can be expressed on OL breakdown rules. For example, the circular convolution $\mathbf{Conv}_n$ is an arity $(2, 1)$ operator. $\mathbf{Conv}_n$ can be performed with the forward and inverse DFTs, which are then recursively broken down with the algorithms like [29, 24].

$$\mathbf{Conv}_n \rightarrow \mathbf{DFT}_{n,-1} \circ P_n \circ (\mathbf{DFT}_{n,1} \times \mathbf{DFT}_{n,1}), \qquad (7)$$

Where $\mathbf{DFT}_{n,-1}$ and $\mathbf{DFT}_{n,1}$ are the inverse and forward transforms, respectively.

## 3 Fast Algorithms

We have learned that the circular convolution is equivalent to the polynomial multiplication in the ring $R[x]/x^n - 1$ and can be done efficiently via the convolution theorem. The linear convolution of size $n$ can be obtained via circular convolution by zero-padding to size $2n$, leading to the polynomial multiplication in $R[x]$. In this section, we first review the Convolution theorem, then present fast algorithms based on the modular FFT[29, 24] and TFT[25, 27].

**Convolution Theorem.** We have learned the equivalence between the linear and circular convolutions and the polynomial multiplication. Next, we introduce the convolution theorem, which leads to fast FFT-based polynomial multiplication algorithms.

The convolution theorem states that the Fourier transform of a convolution is the pointwise product of Fourier transforms. That is, convolution in one domain equals pointwise multiplication in the other domain.

**Theorem 1.** *Let $f$ and $g$ be two polynomials $\in R[x]$ of degree less than $n$, the convolution theorem states that*

$$\mathrm{DFT}(f \circledast g) = \mathrm{DFT}(f) \cdot \mathrm{DFT}(g) \tag{8}$$

*where $\cdot$ denotes pointwise multiplication.*

The forward transforms evaluate $f$ and $g$ at $w^0, \ldots, w^{n-1}$. Its kernel is $x^n - 1$, and the theorem says that DFT mapping $R[x]/x^n - 1 \to R^n$ is a homomorphism of $R$-algebras, where multiplication in $R^n$ is pointwise multiplication of vectors. The following commutative diagram clearly illustrates the mapping relationships:

$$\begin{array}{ccc} (R[x]/x^n - 1)^2 & \xrightarrow{DFT \times DFT} & R^n \times R^n \\ {\scriptstyle circular \atop convolution}\Big\downarrow & & \Big\downarrow{\scriptstyle pointwise \atop multiplication} \\ R[x]/x^n - 1 & \xrightarrow[DFT]{} & R^n \end{array} \tag{9}$$

Then, by applying the inverse transform $\mathrm{DFT}^{-1}$, we can write:

$$f \circledast g = \mathrm{DFT}^{-1}\{DFT(f) \cdot DFT(g)\} \tag{10}$$

By using the aforementioned fast linear transform algorithms, we obtain an efficient algorithm for computing the circular convolution, and thus for polynomial multiplication mod $x^N - 1$. In a ring $R$ that has appropriate roots of unity to support FFT, the convolution in $R[x]/x^n - 1$ and multiplication of polynomials $f, g \in R[x]$ with $\deg(fg) < n$ can now be performed in $O(n \log n)$.

The fast circular convolution algorithm based on the modular DFT can be expressed in the OL as:

$$\mathbf{CirConv}_n \rightarrow \mathbf{ModDFT}_n^{-1} \circ P_n \circ (\mathbf{ModDFT}_n \times \mathbf{ModDFT}_n), \tag{11}$$

where the *cross operator* $\times$ has an arity of $(2, 2)$ in this case and applies the left and right $\mathbf{ModDFT}_n$ to the two input vectors, respectively. Also recall that $\circ$ represents the composition of operations, and $P$ performs the pointwise multiplication with an arity of $(2, 1)$.

Similarly, the linear convolution with zero-padding, can be expressed as:

$$\mathbf{LinConv}_n \rightarrow \mathbf{ModDFT}_{2n}^{-1} \circ P_{2n} \circ ((\mathbf{ModDFT}_{2n} \circ ZP_{2n}) \times (\mathbf{ModDFT}_{2n} \circ ZP_{2n}), \tag{12}$$

where $ZP$ is an operation that pads the input vectors with zeros at the end to the desired length. In practice, the modular DFT based fast linear convolution algorithm (12) requires the input sizes to be powers of two and uses zero padding for arbitrary input sizes.

Other algorithm candidates include by-definition for small convolutions, and factorization of a circular convolution $(\mathbb{Z}[x]/(x^{2n} - 1))$ to a circular convolution $(\mathbb{Z}[x]/(x^n - 1))$ and a nega-circular convolution $(\mathbb{Z}[x]/(x^n + 1))$ as in (13).

$$\mathbb{Z}[x]/(x^{2n} - 1) \cong \mathbb{Z}[x]/(x^n - 1) \times \mathbb{Z}[x]/(x^n + 1) \tag{13}$$

As we can see, the performance of convolutions essentially depends on the performance of the underlying transforms, which justifies our previous focus on the library generation and optimization for the modular FFT and TFT.

Next, we introduce the TFT-based convolution algorithm which achieves reduced arithmetic cost from the underlying TFT and ITFT. Let $R$ be a commutative ring. We assume that $R$ contains a principal $L^{th}$ root of unity $\omega$. Examples of $R$ include $R = \mathbb{Z}/(2^{L/2} + 1)\mathbb{Z}$ where $L = 2^l$ and $\omega = 2$, which appears in the Schönhage-Strassen algorithm for multiplication in $\mathbb{Z}[x]$ [10, 35].

The TFT and ITFT may be used to deduce a polynomial multiplication algorithm in $R[x]$ as follows. Let $g, h \in R[X]$, and $u = gh$. Let $z_1 = 1 + deg(g)$, $z_2 = 1 + deg(h)$, $n = z_1 + z_2 - 1$, and assume that $n \leq L$. Let $g_0, \ldots, g_{z_1-1}$ be the coefficients of $g$ and $h_0, \ldots, h_{z_2-1}$ be the coefficients of $h$. Compute

$$(\hat{g}_0, \ldots, \hat{g}_{n-1}) = \mathbf{TFT}(L, z_1, n) \cdot (g_0, \ldots, g_{z_1-1}),$$
$$(\hat{h}_0, \ldots, \hat{h}_{n-1}) = \mathbf{TFT}(L, z_2, n) \cdot (g_0, \ldots, g_{z_2-1}),$$

and then compute $\hat{u}_i = \hat{g}_i \hat{h}_i$ in $R$ for $0 \leq i < n$. Then $\hat{u}_0, \ldots, \hat{u}_{n-1}$ are the first $n$ Fourier coefficients of $u$, and $u_j = 0$ for all $n \leq j \leq L - 1$ since $n = deg(u) + 1$. Therefore, we recover $u$ via

$$(Lu_0, \ldots, Lu_{n-1}) = \mathbf{ITFT}(L, n, n) \cdot (\hat{u}_0, \ldots, \hat{u}_{n-1}).$$

The TFT-based convolution algorithm can also be expressed in OL as follows:

$$\mathbf{Conv}_n \rightarrow \mathbf{ITFT}_{L,n,n} \circ P_n \circ (\mathbf{TFT}_{L,z_1,n} \times \mathbf{TFT}_{L,z_2,n}). \tag{14}$$

**Computational complexity.** The standard FFT algorithms compute the DFT or inverse DFT using $\frac{L \log L}{2}$ 'butterfly operations'. In contrast, van der Hoeven showed that the TFT and ITFT may be computed using at most $\frac{n \log L}{2} + L$ butterfly operations. Moreover, in the multiplication algorithm sketched above, only $n$ pointwise multiplication are performed, compared to $L$ multiplications incurred by the standard FFT method. Therefore, the ratio of the running time of the TFT/ITFT-based multiplication algorithm over that of the standard FFT-based algorithm is $n/L + O((\log n)^{-1})$, indicating that the performance is relatively smooth as a function of $n$.

## 4 SPIRAL **Enhancements**

SPIRAL lacks the support for base case generation for multiple nonterminals within one library. In the modular polynomial multiplication library, the convolution and the linear transforms each requires a set of base cases for recursion termination and improved performance for the memory hierarchy. Therefore, the *base case generator* module has been extended to take as input and match multiple base case patterns and invoke the corresponding base case generation rules. Fig 2 illustrates the extension, where the dashed lines show the boundaries of the libraries to be generated.

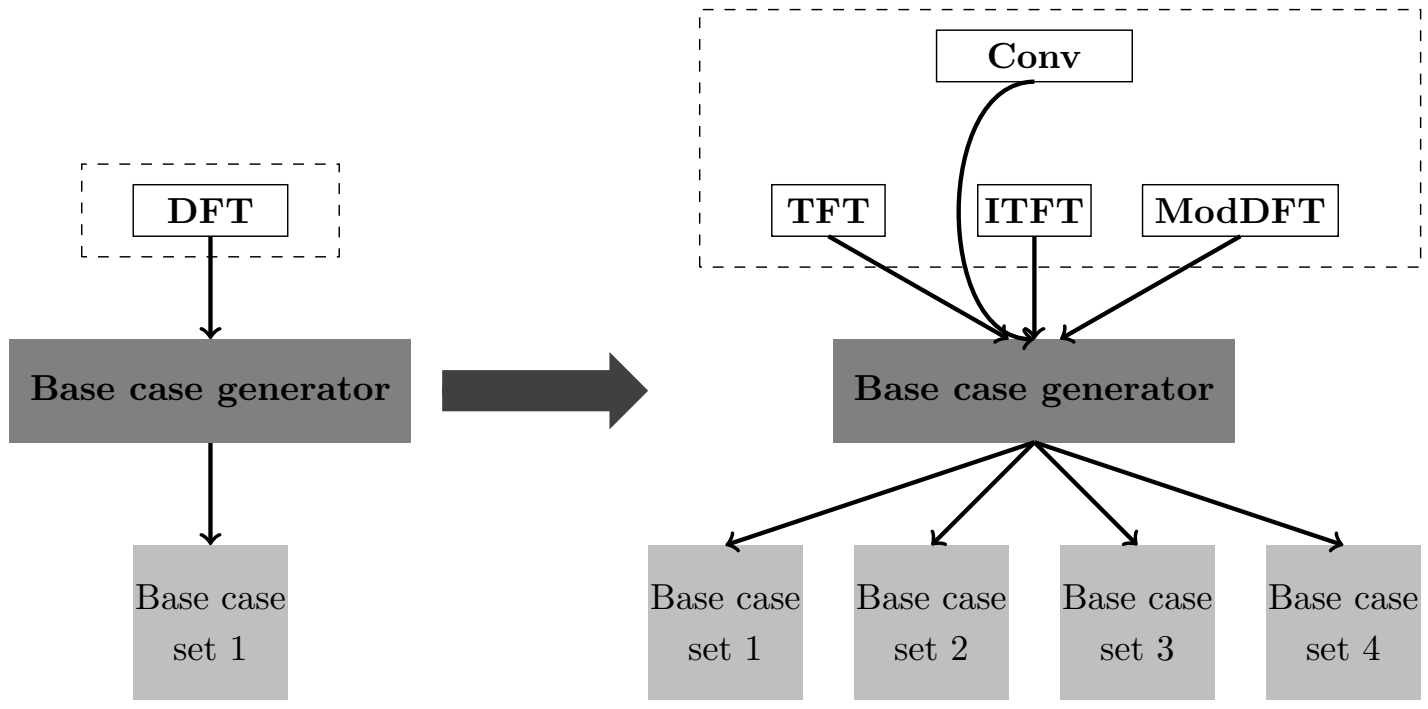

**Fig. 2.** Extension in the base case generator to support multiple base case sets in one library.

The tag propagation is important in the OL formula derivation, especially when substructures are encapsulated in operators such as $\times$, the cross operator. During the development, a small yet severe bug has been discovered which, under certain circumstances, prevented the termination of the recursion step closure computation by generating the so-called "black holes".

For vectorized and multi-threaded library generation, two tags are propagated during the decomposition, namely [*AParLib*, *AVecLib*]. The *AParLib* tag triggers the GT_HPar, and the *AVecLib* tag triggers the GT_HVec, for parallelization and vectorization derivation, respectively. A small rewriting rule named *TTag_Cross* was designed to distribute the tags to the substructures under the cross operator. However, *TTag_Cross* used a left fold operation, which effectively reverses the order of the tags. When the *AParLib* tag is last consumed, it prevents the inclusion of any base cases, resulting in "black holes" in the closure computation which eventually lead to library generation failure. The solution is simple: use the right fold in *TTag_Cross*.

The planning framework has been extended to ensure the mirrored decomposition paths between the permuted TFT and ITFT within the circular convolution library. The DP knowledge entry is extended to include an *execution signature* generated each time when the top-level planning starts. The forward transform first tries to reuse the entry that matches the function signature and the execution signature. If such entry does not exist, it tries to reuse an entry that matches the function signature, and constructs a new entry based on the matched entry and the current execution signature. If no entry is matched on the two said levels, the forward transform creates a DP knowledge entry via the planning framework with its current signature. Then the inverse transform can follow the mirrored decomposition path by reusing the DP knowledge entries for the forward transform with some extra string manipulation.

## 5 Performance Evaluation

This section reports the performance data comparing the TFT-based convolution library and the modular FFT based convolution library, both of which are automatically optimized and generated by SPIRAL. The performance is reported in cycles measured by PAPI [31] and is averaged over 1000 runs with small observed performance variance. All experiments were performed on an Intel Core i7 965 quad-core processor running at 3.2 GHz with 12 GB of RAM. Generated code was compiled with gcc version 4.3.4-1 with optimization set to O3. Vector code used SSE 4.2 with 4-way 32-bit integer vectors.

**Baseline implementation.** An SPIRAL-generated parallel ModDFT-based convolution library using Algorithm (12) is used as the baseline implementation. The reference library pads input to the next power of two, and fully utilizes vector registers and multi-core. Its underlying transform has been shown in [24] that its performance is comparable to optimized fixed-size codes [29], and gains an order-of-magnitude speed-up over hand-optimized library [20]. The ModDFT-based convolution library is represented by the solid black line in Figure 3, which exhibits clear jumps in running time when the lengths cross a power-of-two boundary.

**Autotuned TFT-based convolution library.** The TFT-base convolution library employs the strict general-radix algorithms for base case generation to reduce arithmetic cost. The built-in search engine uses the DP technique that measures the actual running time of smaller transforms as the input to the feedback loop to guide the generation of the base case sizes up to 8. For the library-level recursive breakdown, it applies the relaxed parallel algorithms, which trades off a slightly higher arithmetic cost for vectorization and parallelization. The TFT-based convolution library is represented by the red line in Figure 3

**Speedup.** The 4-thread TFT-based convolution library delivers a speedup of 35% – 41% over the 4-thread high performance ModDFT-based convolution library with length that just crosses a power-of-two boundary. As the length increases, the gap between the TFT-based approach and the full transform based approach decreases.

Overall, the TFT-based convolution library's performance is smooth with respect to the input size. Also note that the relaxation in TFT and ITFT introduces slightly higher arithmetic cost which is bounded by the optimized base case sizes. As a result, mini jumps can be seen between power-of-two jumps, which do not affect the overall smooth performance.

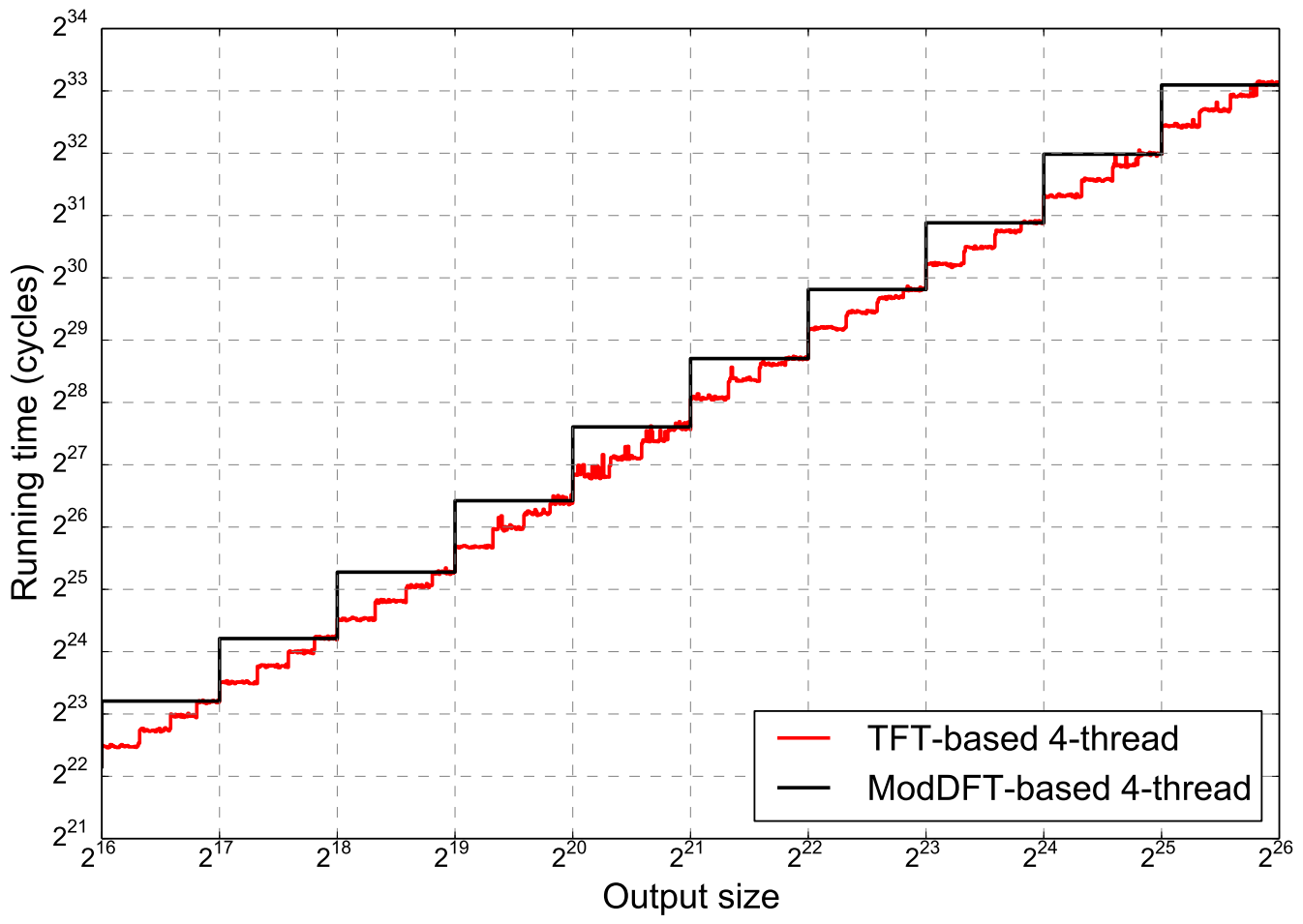

**Fig. 3.** Performance comparison between the SPIRAL-generated TFT-based convolution library and modular DFT-based convolution library

## 6 Conclusion

In this paper we presented the automatic library generation and optimization for modular polynomial multiplication. Then, we presented the convolution theorem which leads to the fast FFT-based convolution algorithms. With the design and implementation of the TFT and ITFT, the FFT-based modular polynomial algorithms can choose from the power of two modular DFT and the TFT. We have shown that the performance gains from previous effort, including the improved modular arithmetic and the parallel modular DFT and TFT, directly contribute to the performance of the automatically generated and optimized library for modular polynomial multiplication. Furthermore, the performance evaluation comparing the modular DFT based library and the TFT-based library shows the practical performance benefits from the new TFT algorithms.